\documentclass{ws-ijmpa}

\newcommand{\be}{\begin{equation}}
\newcommand{\ee}{\end{equation}}
\newcommand{\bea}{\begin{eqnarray}}
\newcommand{\eea}{\end{eqnarray}}
\newcommand{\bean}{\begin{eqnarray*}}
\newcommand{\eean}{\end{eqnarray*}}

\newcommand{\gapproxeq}{\lower
.7ex\hbox{$\;\stackrel{\textstyle >}{\sim}\;$}}
\newcommand{\lapproxeq}{\lower
.7ex\hbox{$\;\stackrel{\textstyle <}{\sim}\;$}}

\begin{document}

\markboth{Q. Zhao}
{Scalar glueball production in $J/\psi$ hadronic decays}

%
\catchline{}{}{}{}{}
%

\title{Scalar glueball production in $J/\psi$ hadronic decays
}

\author{\footnotesize Qiang Zhao}

\address{Institute of High Energy Physics,
CAS,
Beijing, 100049, P.R. China\\
Department of Physics, University of Surrey, Guildford, Surrey GU2 7XH, United Kingdom}

\author{Frank E. Close}

\address{Department of Theoretical Physics,
University of Oxford, Oxford, OX1 3NP, United Kingdom}

\maketitle


\begin{abstract}
The recent data for $J/\psi$ hadronic decays from BES raise ``puzzles"
about the scalar meson properties, which are different
from the naive OZI rule expectation. In this work, we study
the implications of a glueball-$Q\bar{Q}$
mixing scheme in the scalar meson productions in $J/\psi$ hadronic decays.
We show that the significant OZI violations in
$J/\psi\to V f_0^i$ ($i=1,2,3$ correspond to $f_0(1710)$, $f_0(1500)$ and $f_0(1370)$, respectively)
are consistent with the requirement of strong glueball-$Q\bar{Q}$
mixings within the scalar meson wavefunctions.

\keywords{Scalar glueball; $J/\psi$ hadronic decay; OZI rule.}
\end{abstract}


The recent data for $J/\psi$ hadronic decays from BES raised new ``puzzles"
about the scalar meson ($f_0(1710)$, $f_0(1500)$ and $f_0(1370)$) properties~\cite{bes-phi,bes-plb}.
These states were found to behave contrary to expectations
based on naive application of the Okubo-Zweig-Iizuka (OZI) rule
in $J/\psi\to \phi f_0^i\to \phi PP$ and $J/\psi\to \omega f_0^i\to \omega PP$,
where $PP$ denotes pseudoscalar meson pairs. The puzzles are outlined as follows:
i) the $f_0(1370)$ has been seen clearly in $J/\psi\to \phi\pi\pi$,
but not in $J/\psi\to \omega\pi\pi$;
ii) there is no peak of the $f_0(1500)$ directly seen in
$J/\psi\to \phi K\bar{K}, \ \omega K\bar{K}, \ \phi\pi\pi, \ \omega\pi\pi$,
though its production in proton-proton scattering is quite clear~\cite{wa102};
iii) the $f_0(1710)$ is observed clearly in both $J/\psi\to \phi K\bar{K}$ and
$J/\psi\to \omega K\bar{K}$, but with
$br_{J/\psi\to \omega f_0(1710)\to \omega K\bar{K}}/br_{J/\psi\to\phi f_0(1710)\to\phi K\bar{K}}\simeq 
6$,
which is against a simple $s\bar{s}$ configuration
for this state;
iv) another scalar $f_0(1790)$, which is seen in $\pi\pi$ rather than $K\bar{K}$, is also reported to 
be
distinguished from the $f_0(1710)$.

We will present an anlysis of $J/\psi\to V f_0^i$, where $V=\phi$ or $\omega$.
Based on a glueball-$Q\bar{Q}$ mixing scheme proposed by Close {\it et 
al.}~\cite{close-amsler,close-kirk},
we will show that the scalar production in $J/\psi\to V f_0^i$ is sensitive to
the scalar meson structures. In particular, the strong glueball-$Q\bar{Q}$ mixing
implies a significant OZI-rule violation in the reaction. This will not only  explain the ``scalar 
puzzles"
outlined above, but also predict detectable effects in other production channels.

In the basis of $|G\rangle =| gg\rangle$,
$|s\bar{s}\rangle $ and
$|n\bar{n}\rangle \equiv |u\bar{u}+d\bar{d}\rangle/\sqrt{2}$,
the scalars, $f_0^{1,2,3}$, can be expressed as a mixture of
$G$, $s\bar{s}$ and $n\bar{n}$ via a perturbative potential transition,
$ f\equiv \langle s\bar{s} | \hat{V} | G\rangle
=\langle n\bar{n} | \hat{V} | G\rangle /\sqrt{2}$,
where $f$ is the flavour independent mixing strength.
We can then express these $f_0$ states as
\be
\left(
\begin{array}{c}
|f_0(1710)\rangle \\
|f_0(1500)\rangle \\
|f_0(1370)\rangle
\end{array}
\right) =
\left(
\begin{array}{ccc}
x_1 & y_1 & z_1 \\
x_2 & y_2 & z_2 \\
x_3 & y_3 & z_3
\end{array}
\right)
\left(
\begin{array}{c}
|G\rangle \\
|{s\bar{s}}\rangle \\
|{n\bar{n}}\rangle
\end{array}
\right) \ ,
\ee
where $x_i$, $y_i$ and $z_i$ are the mixing matrix elements determined
by the perturbative transitions~\cite{close-amsler,close-kirk}, and satisfy
both unitary and orthogonal conditions. The factorizations for $f_0^i\to PP$
are given in Ref.~\cite{close-zhao-f0}.

To be consistent with the decay of the $f_0$ states,
we factorize their production in $J/\psi\to V f_0^i$ by defining the
transitions:
\bea
M_{J/\psi\to\phi f_0^i} & = &
\langle f_0^i|G\rangle\langle G|V_\phi|J/\psi\rangle
+ \langle f_0^i|s\bar{s}\rangle\langle s\bar{s}|V_\phi|J/\psi\rangle
+ \langle f_0^i|n\bar{n}\rangle\langle n\bar{n}|V_\phi|J/\psi\rangle \nonumber\\
&=& x_i M_{\phi G} + y_i M_{\phi (s\bar{s})}
+ z_i M_{\phi(n\bar{n})} \ ,
\eea
where $V_\phi$ is the potential for producing the $\phi$ meson recoiling against the
scalar $f_0$. Note that the production of the glueball component
is at the same nominal order as the singly OZI disconnected (SOZI) processes
of producing an $s\bar{s}$ component. The production of the non-strange
$n\bar{n}\equiv (u\bar{n}+d\bar{d})/\sqrt{2}$ recoiling the $\phi$ meson will
be doubly OZI suppressed (DOZI).
The implication of such a factorization for the scalar productions
in $J/\psi\to V f_0$ is that significant OZI-rule violation would be
expected. This can be examined by comparing the naive OZI expectations
with the experimental data. Similar to $J/\psi\to \phi f_0^i$, we can also make the
factorization for $J/\psi\to \omega f_0^i$, where the DOZI process will
be the production of $\omega$ recoiling the isoscalar $s\bar{s}$ component.
In both cases, the vector meson $\omega$ and $\phi$ serve
as a flavor filter for the quark components of the scalar mesons
in the limit of naive OZI rule.

For $J/\psi\to \phi f_0^i$, amplitudes $M_{\phi G}$ and $M_{\phi (s\bar{s})}$
are of the same nominal order in perturbative QCD.
Thus, we assume $M_{\phi G}\simeq M_{\phi (s\bar{s})}$.
Transition $M_{\phi(n\bar{n})}$ can only occur via the DOZI process, which
is of $O(\alpha_s)$ relative to the SOZI ones in the perturbative regime.
However, due to the complexity of the hadronization
of gluons into $Q\bar{Q}$, the role played by the DOZI
processes is still not well-understood.
The study of Refs.~\cite{isgur-geiger,lipkin-zou} gives some hints of
the presence of strong OZI-rule violations within the scalars,
which suggests that the DOZI processes may be expected to occur at similar
strength to the SOZI ones.
To estimate their contributions,
we introduce another parameter $r$,
which describes the relative strength between the DOZI
and SOZI processes such that:
\be
\label{fact-amp}
\langle G|V_\phi |J/\psi\rangle
\simeq \langle s\bar{s}|V_\phi|J/\psi\rangle
\simeq \frac{1}{\sqrt{2}r}\langle n\bar{n}|V_\phi|J/\psi\rangle \ ,
\ee
where $\sqrt{2}$ is from the normalization of $n\bar{n}$, and
we have eventually assumed that the potential
is OZI-selecting. The left equation holds since
$\langle G|V_\phi |J/\psi\rangle$ and
$\langle s\bar{s}|V_\phi|J/\psi\rangle $ both are of the same
nominal order and both are SOZI transitions.
The right equation then distinguishes the DOZI
process from the SOZI ones.

The above assumption will then allow us to express
the partial decay width as
\bea
\label{decay-1}
\Gamma_{J/\psi\to\phi f_0^i\to\phi PP} & = &
\frac{|{\bf p}_{\phi i}|}{|{\bf p}_{\phi G}|}
br_{f_0^i\to PP} [  x_i + y_i  + \sqrt{2}r z_i ]^2
\Gamma_{J/\psi\to \phi G} \ ,
\eea
where ${\bf p}_{\phi i}$ and ${\bf p}_{\phi G}$ are the momenta of
the $\phi$ meson in  $J/\psi\to \phi f_0^i$ and the virtual $J/\psi\to \phi G$,
respectively. Their ratio gives the kinematic correction
for the decays of $J/\psi$
to the $\phi$ meson and states with different masses, i.e.
$M_i\neq M_G$.
$\Gamma_{J/\psi\to \phi G}$
is the $J/\psi$ decay width into $\phi$ and glueball $G$.

Similarly, for $J/\psi\to\omega f_0^i\to \omega PP$,
we have the partial decay width
\bea
\label{decay-2}
\Gamma_{J/\psi\to\omega f_0^i\to\omega PP} & = &
\frac{|{\bf p}_{\omega i}|}{|{\bf p}_{\omega G}|}
br_{f_0^i\to PP} [ x_i + r y_i + \sqrt{2} z_i ]^2
\Gamma_{J/\psi\to \omega G} \ ,
\eea
where $\Gamma_{J/\psi\to \omega G}$
is the $J/\psi$ decay
width into $\omega$ and pure glueball $G$; ${\bf p}_{\omega i}$
and ${\bf p}_{\omega G}$ are the three momenta of $\omega$ 
in $J/\psi\to\omega f_0^i$ and $J/\psi\to \omega G$, respectively.

The flavour-blind assumption implies $2 \Gamma_{J/\psi\to \phi G}/|{\bf p}_{\phi G}|
=\Gamma_{J/\psi\to \omega G}/|{\bf p}_{\omega G}| $.
For given $f_0^i$, this relation will allow us to relate the $\omega$ and $\phi$ channel
together, and then the parameter $r$ can be determined by the experimental data
for the partial decay widths. For example, the BES data,
$\Gamma_{J/\psi\to\phi f_0^1\to\phi K\bar{K}}/\Gamma_{J/\psi}^{T}
=(2.0\pm 0.7 )\times 10^{-4}$ and
$ \Gamma_{J/\psi\to\omega f_0^1\to\omega K\bar{K}}/\Gamma_{J/\psi}^{T}
=(13.2\pm 2.6) \times 10^{-4}$, allow us to derive
\be
\label{ddd}
r=\frac{1}{\chi_0 y_1-\sqrt{2} z_1}
\left[(1-\chi_0)x_1 +y_1 -\sqrt{2}\chi_0 z_1\right] \ ,
\ee
by taking the ratio
$R^{OZI}_i={\Gamma_{J/\psi\to\phi f_0^i\to\phi K\bar{K}}}/
{\Gamma_{J/\psi\to\omega f_0^i\to\omega K\bar{K}}} $,
with $i=1$ (for $f_0(1710)$) and $\chi_0\equiv (2R^{OZI}_1 |{\bf p}_{\omega 1}|/|{\bf p}_{\phi 1} 
|)^{1/2}$.

The WA102 data for $f_0^i\to PP$~\cite{wa102} allow us to determine the parameters for
the glueball-$Q\bar{Q}$ mixing matrix elements~\cite{close-zhao-f0}.
In association with the BES data~\cite{bes-phi,bes-plb},
$br(f_0(1710)\to \pi\pi)/br(f_0(1710)\to K\bar{K})< 0.11$ with 95\% confidence level,
we find $M_{s\bar{s}}>M_G>M_{n\bar{n}}$ is favored and it leads to
\bea
\label{mix-2}
|f_0(1710)\rangle & = & 0.36 |G\rangle + 0.93 |{s\bar{s}}\rangle + 0.09 |{n\bar{n}}\rangle \nonumber\\
|f_0(1500)\rangle & = & -0.84|G\rangle + 0.35 |{s\bar{s}}\rangle -0.41 |{n\bar{n}}\rangle \nonumber\\
|f_0(1370)\rangle & = & 0.40|G\rangle - 0.07 |{s\bar{s}}\rangle -0.91 |{n\bar{n}}\rangle  \ ,
\eea
where the relative signs and magnitudes among $G$, $s\bar{s}$ and $n\bar{n}$ turn out to be a stable
pattern of the wavefunction mixings. With the mixing matrix elements, we derive $r=2.2$,
which suggests significant OZI-rule violations in $J/\psi\to Vf_0^1$. This property
should be valid for $f_0^2$ and $f_0^3$ in their productions, and consistent with the basic
assumption of the strong glueball-$Q\bar{Q}$ mixings within the scalars.
Furthermore, it can be examined by comparing the predictions for $J/\psi\to V f_0^{2,3}$
with the data.

This factorization scheme provides an estimate of the pure
glueball production rates, $br_{J/\psi\to\phi G}=1.63\times
10^{-4}$ and $br_{J/\psi\to\omega G}=3.75\times 10^{-4}$. Although
the existence of a pure glueball is not clear, these numbers can be
compared with the lattice QCD results in the future. The magnitude
of the production rates also suggests possible large DOZI
contributions.

From Eqs.~\ref{decay-1} and \ref{decay-2}, we calculate the
branching ratios, $br_{J/\psi\to Vf_0^i\to VPP}$, for $f_0^{2,3}$
in comparison with the BES data. Without the contributions from
the DOZI processes, i.e. $r=0$, we cannot reproduce the pattern of
the data as described at the beginning. Taking into account the
DOZI contributions, we obtain consistent results, which naturally
explain the puzzles without making dramatic assumptions. The
results are presented in Table~\ref{tab-1}. It shows that for the
$f_0(1500)$ and $f_0(1370)$, their productions in $J/\psi\to
\omega K\bar{K}$ are indeed much smaller than in $\phi K\bar{K}$.
In contrast, these two states have larger production rates in
$J/\psi\to \phi \pi\pi$ than in $\omega \pi\pi$.

\begin{table}[h]
\tbl{Predictions for $br_{J/\psi\to V f_0^i\to  V PP}$ in comparison
with the BES data (in the round brackets). The numbers are in unit of
$10^{-4}$. The symbol `$\cdots$' means
that signals of the corresponding states are absent from experiment.}
{\begin{tabular}{@{}ccccccccc@{}} \toprule
\hline
Scalars & $K\bar{K} \ (\phi)$  & Data
& $K\bar{K} \ (\omega)$  &  Data & $\pi\pi \ (\phi)$ & Data & $\pi\pi \ (\omega)$ & \\[1ex]
\hline
$f_0(1710)$ & {\bf 2.00 } & (2.0$\pm$ 0.7)
& {\bf 13.20 } & (13.2$\pm$ 2.6) & 0.22 & $\cdots$  & 1.45
& $\cdots$  \\[1ex]
$f_0(1500)$ & 0.46 & (0.8$\pm$ 0.5)
& 0.13 & $\cdots$ & 1.89 &  (1.7$\pm$ 0.8) & 0.52 & $\cdots$ \\[1ex]
$f_0(1370)$ & 0.26 & (0.3$\pm$ 0.3)
& 0.09 & $\cdots$   & 2.63 & (4.3$\pm$ 1.1) & 0.94 & $\cdots$ \\[1ex]
\hline
\label{tab-1}
\end{tabular}}
\end{table}

To summarize, the recent data from BES reveal a very rich source of strong QCD dynamics
in the scalar meson sector. We show that strong glueball-$Q\bar{Q}$ mixings
imply the strong OZI-rule violations in $J/\psi$ hadronic decays.
It may partially explain why signals for the pure scalar glueball have so far remained hidden.
Interestingly, we find that the mixing matrix exhibits stable features in both relative phases and
wavefunction densities, to which their productions in $J/\psi\to V f_0^i$ are sensitive.
Based on such a mixing scheme, it is then natural to understand
the distinguished charactor of $f_0(1790)$ from the $f_0(1710)$,
for which the former can be interpreted as the radial excited state of the $f_0(1370)$.
Further tests of such a glueball-$Q\bar{Q}$ mixing scheme
in scalar meson productions~\cite{close-ichep04,zhao-zou-ma,zhao-chic} are still needed
to provide more evidence for the existence of the scalar glueball.

This work is supported, in part,
by grants from the U.K. EPSRC Advanced Fellowship (Grant No. GR/S99433/01),
and the Institute of High Energy Physics, Chinese Academy of Sciences.

\end{document}